\documentclass[aps,twocolumn,english,superscriptaddress,citeautoscript,preprintnumbers,amsmath,amssymb,floatfix,footinbib]{revtex4-1}
\usepackage{hyperref}
\usepackage{xcolor}
\hypersetup{colorlinks=true,linkcolor=red,citecolor=blue}
\usepackage{amsmath}
\usepackage{graphicx}
\usepackage{dcolumn}
\usepackage{float}

\begin{document}
\title{Colossal negative magnetoresistance in the complex charge density wave regime of an antiferromagnetic Dirac semimetal}

\author{Ratnadwip Singha}
    \affiliation{Department of Chemistry, Princeton University, Princeton, New Jersey 08544, USA}

\author{Kirstine J. Dalgaard}
    \affiliation{Department of Chemistry, Princeton University, Princeton, New Jersey 08544, USA}

\author{Dmitry Marchenko}
    \affiliation{Helmholtz-Zentrum Berlin f\"{u}r Materialien und Energie Elektronenspeicherring BESSY II, Albert-Einstein-Stra{\ss}e 15, 12489 Berlin, Germany}

\author{Maxim Krivenkov}
    \affiliation{Helmholtz-Zentrum Berlin f\"{u}r Materialien und Energie Elektronenspeicherring BESSY II, Albert-Einstein-Stra{\ss}e 15, 12489 Berlin, Germany}

\author{Emile D. L. Rienks}
    \affiliation{Helmholtz-Zentrum Berlin f\"{u}r Materialien und Energie Elektronenspeicherring BESSY II, Albert-Einstein-Stra{\ss}e 15, 12489 Berlin, Germany}

\author{Milena Jovanovic}
    \affiliation{Department of Chemistry, Princeton University, Princeton, New Jersey 08544, USA}

\author{Samuel M. L. Teicher}
    \affiliation{Materials Department and Materials Research Laboratory, University of California Santa Barbara, CA 93106, USA}

\author{Jiayi Hu}
    \affiliation{Department of Physics, Princeton University, Princeton, New Jersey 08544, USA}

\author{Tyger H. Salters}
    \affiliation{Department of Chemistry, Princeton University, Princeton, New Jersey 08544, USA}

\author{Jingjing Lin}
    \affiliation{Department of Physics, Princeton University, Princeton, New Jersey 08544, USA}

\author{Andrei Varykhalov}
    \affiliation{Helmholtz-Zentrum Berlin f\"{u}r Materialien und Energie Elektronenspeicherring BESSY II, Albert-Einstein-Stra{\ss}e 15, 12489 Berlin, Germany}

\author{N. Phuan Ong}
    \affiliation{Department of Physics, Princeton University, Princeton, New Jersey 08544, USA}

\author{Leslie M. Schoop}
    \affiliation{Department of Chemistry, Princeton University, Princeton, New Jersey 08544, USA}

\begin{abstract}
Colossal magnetoresistance (MR) is a well-known phenomenon, notably observed in hole-doped ferromagnetic manganites. It remains a major research topic due to its potential in technological applications. Though topological semimetals also show large MR, its origin and nature are completely different. Here, we show that in the highly electron doped region, the Dirac semimetal CeSbTe demonstrates similar properties as the manganites. CeSb$_{0.11}$Te$_{1.90}$ hosts multiple charge density wave (CDW) modulation-vectors and has a complex magnetic phase diagram. We confirm that this compound is an antiferromagnetic Dirac semimetal. Despite having a metallic Fermi surface, the electronic transport properties are semiconductor-like and deviate from known theoretical models. An external magnetic field induces a semiconductor-metal-like transition, which results in a colossal negative MR. Moreover, signatures of the coupling between the CDW and a spin modulation are observed in resistivity. This spin modulation also produces a giant anomalous Hall response.
\end{abstract}

\maketitle

Magnetoresistance (MR), i.e., a change (both increase and decrease) in the electrical resistivity of a material with application of a magnetic field is an extensively studied phenomenon in condensed matter physics. Though known for decades, it remains relevant to this day due to the wide range of technological applications \cite{Dieny,Parkin1,Asamitsu,Bibes,Freitas} as well as open questions about its origin \cite{Lee1982,Camley,Millis,Abrikosov} in a large variety of systems. Depending on the observed features, types of material, and physical mechanisms involved, different terms have been introduced to categorize a large MR. For example, the giant MR (GMR) appears in ferromagnetic-non-magnetic multilayer heterostructures, where a small magnetic field ($\sim$few tenths of a tesla) results in electronic conduction between the spin-polarized layers and hence a sharp drop in resistivity \cite{Dieny,Parkin1,Baibich,Binasch}. On the other hand, large negative colossal MR (CMR) is observed near the Curie temperature of hole-doped manganite perovskites, pyrochlores, and spinel compounds as a result of the transition from a paramagnetic insulating to a ferromagnetic metallic state \cite{Ramirez1,Tokura,Jin,Shimakawa,Ramirez2}. However, the origin of CMR is different in each material family. Among these, manganites are the most explored, where primarily the double-exchange interaction between mixed valence Mn ions leads to the metallic ferromagnetic state \cite{Zener,Anderson,Gennes}. In recent years, a very large positive MR has been reported in topological semimetals (TSMs) owing to a combination of relativistic quasiparticle excitations, high-mobility charge carriers, and highly anisotropic Fermi surface properties \cite{Ali,Liang,Shekhar,Tafti,Singha1}. In a special experimental configuration (collinear electric and magnetic fields), a weak negative MR can also be obtained in TSMs due to the relativistic chiral anomaly \cite{Xiong,Huang}.

\begin{figure*}
\includegraphics[width=0.70\textwidth]{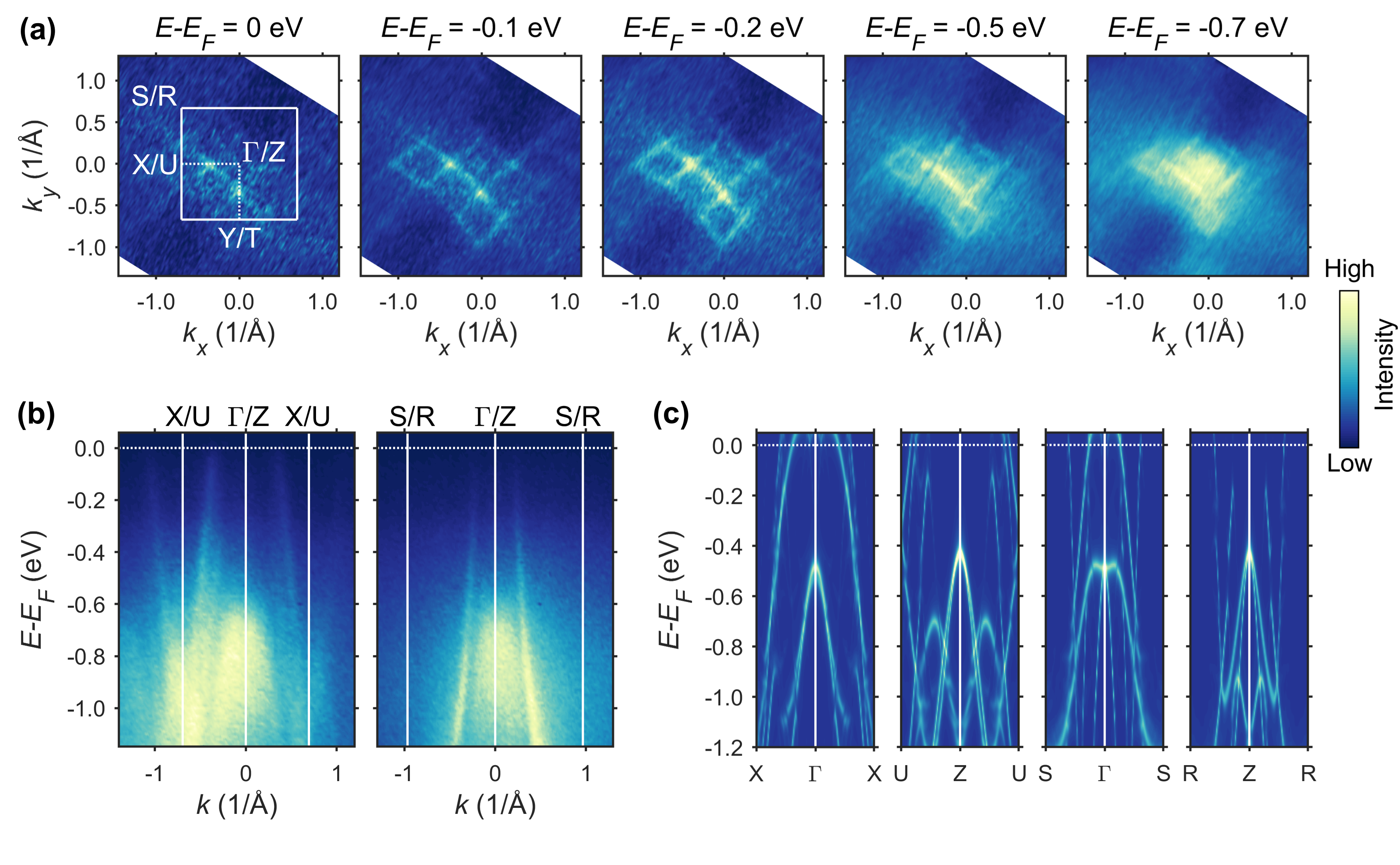}
\caption{ARPES spectra for CeSb$_{0.11}$Te$_{1.90}$ at 1.5 K measured with a photon energy of 110 eV. (a) Constant energy contours at different energy ($E$) values from the Fermi energy ($E_{F}$) to $E_{F}-$0.7 eV. (b) Experimental and (c) theoretical electronic band dispersion along the high symmetry directions $X$($U$)-$\Gamma$($Z$)-$X$($U$) and $S$($R$)-$\Gamma$($Z$)-$S$($R$).}
\end{figure*}

TSMs host linearly dispersing bands in their electronic band structure which are protected by crystallographic symmetries along with lattice inversion and/or time-reversal symmetry \cite{Liu1,Liu2,Lv,Xu}. Hence, they provide an opportunity to study the relativistic particle dynamics in low-energy systems. Through first-principles calculations, a large number of TSMs has been identified over the last few years \cite{Bradlyn,Vergniory1,Zhang,Vergniory2}. Square-net materials are candidates that show probably the cleanest signatures of non-trivial topological bands \cite{Schoop1,Muechler,Singha2,Schoop2,Hosen,Yue,Regmi}. These compounds possess a two-dimensional square-net atomic motif in the crystal structure, which produces isolated Dirac cones at the Fermi energy ($E_{F}$) in the electronic band structure \cite{Klemenz} with the largest reported energy range of linear dispersion \cite{Schoop1}. $Ln$SbTe ($Ln$=lanthanides), which contain magnetic rare-earth elements, are a subgroup of the square-net family, and they make up one of the rare examples of time-reversal symmetry broken magnetic Dirac semimetals \cite{Schoop2,Hosen,Yue,Regmi,Pandey}. Furthermore, by changing the number of electrons per atom at the Sb square-net via chemical substitution, structural distortions and charge density wave (CDW) can be induced in $Ln$SbTe \cite{DiMasi,Lei1,Singha3,Li2}. While electron filling moves the Fermi energy, thus offering access to new bands in the electronic transport experiments, CDWs open gaps at the Fermi surface and yield new quantum states. It has been shown that CDWs clean the band structure and create an ideal non-symmorphic Dirac semimetal state in electron doped GdSbTe \cite{Lei2} as well as lead to rich magnetism and a potential skyrmionic phase \cite{Lei3}.

Here, we focus on the antiferromagnetic (AFM) Dirac semimetal CeSbTe \cite{Schoop2}. With application of an external magnetic field, the band structure of CeSbTe can be tuned to realize Weyl and higher order topological states \cite{Schoop2}. Recently, by substituting Sb with Te on the square-net site, the evolution of the CDW has been investigated in CeSb$_{x}$Te$_{2-x-\delta}$ ($\delta$ is the vacancy concentration in the crystal) \cite{Singha3}. As illustrated in Fig. S1(a), a CDW appears at around $x<$0.79 accompanied by a structural distortion from the tetragonal to an orthorhombic phase. The associated modulation wave-vector (\textit{\textbf{q}}) changes continuously as a function of $x$. At the highest electron filling range 0.10$\leq$$x$$<$0.34, a complex CDW ordering is observed, represented by multiple \textit{\textbf{q}}-vectors [Fig. S1(a)]. The CDW also modifies the AFM ground state. Especially in the region of multiple \textit{\textbf{q}}-vectors, a magnetic field-induced ``devil's staircase'' ordering is reported in the magnetization [Fig. S1(b)] \cite{Singha3}. A series of fractionally quantized magnetization plateaus originate from the coupling between the CDW and a spin modulation along the $c$-axis through Ruderman-Kittel-Kasuya-Yosida (RKKY) interaction. First-principles calculations suggest that a rigid band model can be assumed where the electron filling moves $E_{F}$, and the CDW gaps out a number of band crossings. At $x$=0.11, several Dirac nodes persist at or near $E_{F}$ \cite{Singha3}.

In this letter, we combine angle-resolved photo-emission spectroscopy (ARPES) and electronic transport measurements to unveil the interplay between the CDW, spin modulation, and topological states in the multiple \textit{\textbf{q}}-vector regime of CeSb$_{x}$Te$_{2-x-\delta}$. From ARPES results, we confirm that CeSb$_{0.11}$Te$_{1.90}$ is a Dirac semimetal. In spite of having a non-zero density of states (DOS) at $E_{F}$, this material shows a semiconductor-like behavior, which does not fit into conventional theoretical models for thermally activated transport or localization effects. An external magnetic field instigates a semiconductor-to-metal-like transition and, as a result, a colossal negative MR. Moreover, prominent signatures of the coupling between the CDW and spin modulation are observed in the MR data. In the low magnetic field region, CeSb$_{0.11}$Te$_{1.90}$ also shows a giant anomalous Hall effect, which might emerge from the collective spin-excitation.

Single crystals of CeSb$_{0.11}$Te$_{1.90}$ were grown with chemical vapor transport using iodine as the transport agent. The obtained crystals were characterized by powder/single crystal x-ray diffraction measurements and the chemical compositions were determined by energy-dispersive x-ray spectroscopy (EDX). Additional details about the crystal growth and characterization can be found in our earlier report \cite{Singha3}. ARPES experiments were performed on in-situ cleaved crystals in ultrahigh vacuum ($\sim$10$^{-10}$ mbar). The spectra were recorded using the one-cube ARPES set-up installed at the UE112-PGM2b beam-line at the BESSY-II synchrotron, with various photon energies ($h\nu$) ranging from 40 to 130 eV. Density functional theory (DFT) calculations were performed in VASP v5.4.4 \cite{kresse1994abinitio,kresse1996efficient,kresse1996efficiency} using the PBE functional \cite{perdew1996generalized}. Similar to CeSbTe \cite{Schoop2}, localization of the Ce $f$-orbitals was corrected using a Hubbard potential of $U$=6 eV \cite{dudarev1998electron}. PAW potentials \cite{blochl1994projector,kresse1999from} were selected based on the v5.4.4 recommendations. Calculation for CeSb$_{0.11}$Te$_{1.90}$ were performed on the 3$\times$3$\times$2 supercell of Ce$_{36}$Te$_{68}$Sb$_{4}$ with Fermi levels adjusted based on the electron counts obtained from the experimental data. A plane wave energy cutoff of 400 eV and a $k$-mesh density, $l$=30 (corresponding to 2$\times$2$\times$2 $\Gamma$-centered $k$-meshes for CeSb$_{0.11}$Te$_{1.90}$ supercell) were used. Unfolded spectral functions for the supercells in the subcell BZ were calculated following the method of Popescu and Zunger \cite{popescu2012extracting} in VaspBandUnfolding. The electronic transport experiments were performed in a physical property measurement system (Quantum Design), equipped with a sample stage rotator using the ac transport option. Pre-patterned electrodes in six-probe geometry for resistivity and Hall measurements, were deposited directly on the single crystals by gold evaporation. Gold wires were attached to these electrodes using conducting silver paste.

\begin{figure}
\includegraphics[width=0.5\textwidth]{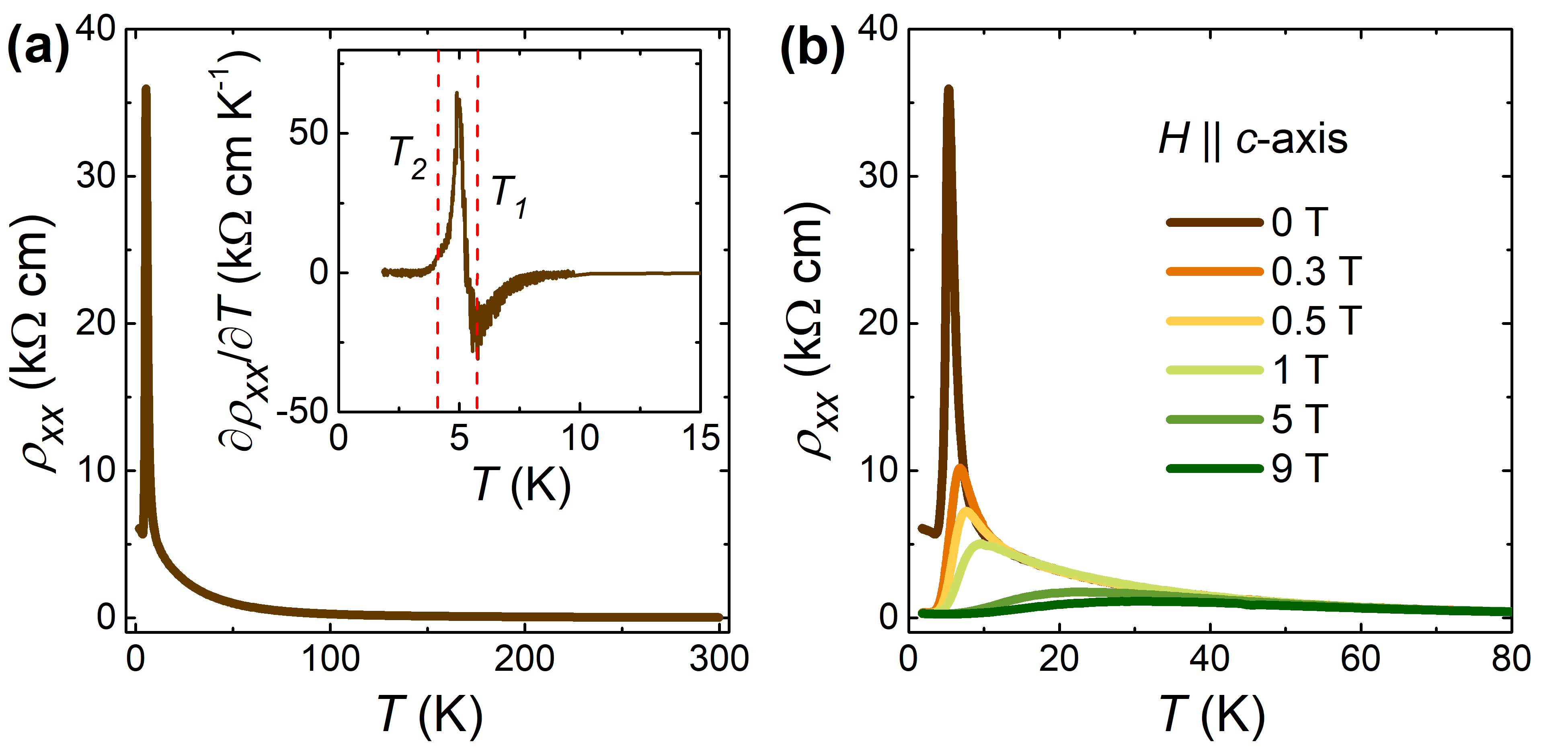}
\caption{Temperature dependent electronic transport properties of CeSb$_{0.11}$Te$_{1.90}$. (a) Temperature ($T$) dependence of the resistivity ($\rho_{xx}$). Inset shows the first order derivative of $\rho_{xx}$($T$), revealing two transition temperatures (dashed vertical lines). (b) $\rho_{xx}$($T$) curves when an external magnetic field is applied along the crystallographic $c$-axis.}
\end{figure}

In Fig. 1, we present the results of the ARPES measurements, measured with a photon energy of 110 eV (see Fig. S2 for an additional photon energy). The experiments were performed at 1.5 K, which is below the reported AFM N{\'e}el temperature ($T_{N}$) 4.27 K for CeSb$_{0.11}$Te$_{1.90}$ \cite{Singha3}. In Fig. 1(a), we show the constant energy contours at different energy ($E$) values from $E_{F}$ to $E$=$E_{F}$$-$0.7 eV. Similar to CeSbTe \cite{Schoop2}, a diamond shaped Fermi surface is observed. Unlike the parent compound, Fermi surface nesting and corresponding energy gap opening is expected in CeSb$_{x}$Te$_{2-x-\delta}$ for x$<$0.79 due to the presence of the CDW. However, despite the strong CDW modulation observed in CeSb$_{0.11}$Te$_{1.90}$ in single crystal x-ray diffraction \cite{Singha3}, the band structure remains rather metallic. The previously reported theoretical band structure for CeSb$_{0.11}$Te$_{1.90}$ showed that several Dirac cones cross the Fermi energy with the nodes residing above $E_{F}$ \cite{Singha3}. Along the $\Gamma$-$X$ and $\Gamma$-$S$ directions the Dirac points are near $E_{F}$. In Figs. 1(b) and 1(c), the measured band dispersions and the calculated band structure, respectively, are plotted along the high symmetry directions $X$-$\Gamma$-$X$ and $S$-$\Gamma$-$S$. For both directions, clean linearly dispersing bands are observed, which agree well with the theoretical results. These bands can be clearly resolved down to $E$=$E_{F}$$-$1.0 eV, therefore confirming the robustness of the Dirac semimetal state in CeSb$_{0.11}$Te$_{1.90}$. To provide an easier comparison, we have overlaid the calculated band structure with the ARPES spectra in Fig. S3. The ARPES spectra measured with photon energy of 70 eV at $\sim$30 K (Fig. S2) also shows similar features, thus indicating that the antiferromagnetic ordering does not modify the topological electronic state.

Despite the absence of any apparent energy gap at the Fermi surface, the resistivity ($\rho_{xx}$) for CeSb$_{0.11}$Te$_{1.90}$ shows a semiconductor-like temperature ($T$) dependence [Fig. 2(a)] when the current is applied along the $b$-axis. From 300 K, $\rho_{xx}$ increases monotonically with decreasing temperature, followed by a sharp peak and sudden drop in the low-temperature region. The first order derivative of $\rho_{xx}$($T$) in the inset of Fig. 2(a) reveals two transition temperatures, $T_{1}$$\sim$5.8 K and $T_{2}$$\sim$4 K. Although the change in the slope near $T_{2}$ is subtle, a prominent signature is observed for another crystal (sample 2), from the same batch, as plotted in Fig. S4. At this electron filling for CeSb$_{x}$Te$_{2-x-\delta}$, a short range ferromagnetic (FM) ordering is reported to coexist with the AFM ground state \cite{Singha3}. The ordering temperature ($T_{C}$) of this FM state is just below $T_{N}$. By comparing the resistivity with the previously reported magnetization data \cite{Singha3}, we conclude that $T_{1}$ corresponds to the AFM transition, whereas $T_{2}$ represents $T_{C}$. Both of these transition temperatures are a bit higher as compared to those obtained from magnetization measurements, which might be related to the difference in applied magnetic field between the two measurements. Upon the application of an external magnetic field ($H$) along the crystallographic $c$-axis, the $\rho_{xx}$($T$) curve undergoes a dramatic change [Fig. 2(b)]. With a field of just 0.3 T, the resistivity drops to one third of the maximum value at $T_{1}$ and the peak shifts to higher temperature. This trend continues with increasing field strength and a semiconductor-metal-like transition becomes apparent in the low-temperature region. Even at the highest field of 9 T, a broad maximum persists at around 30 K. The observed features suggest that a very high magnetic field might be needed to completely suppress the semiconductor-like behavior. Similar properties are also obtained for sample 2 (Fig. S4). From the reported magnetization data, we know that for CeSb$_{0.11}$Te$_{1.90}$, a fully spin-polarized state can be achieved at a small field of $\sim$0.2 T \cite{Singha3}. Therefore, magnetic ordering alone cannot explain the low-temperature behavior of $\rho_{xx}$.

\begin{figure*}
\includegraphics[width=0.80\textwidth]{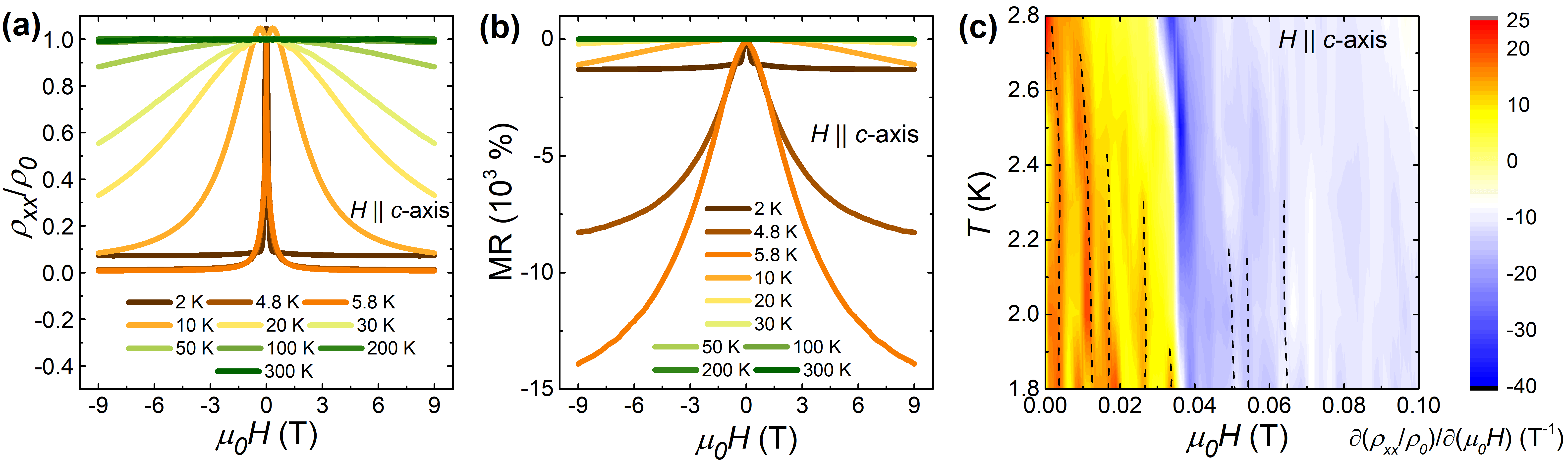}
\caption{Magnetic field dependent electronic transport properties of CeSb$_{0.11}$Te$_{1.90}$. (a) Magnetic field ($H$) dependence of the normalized resistivity ($\rho_{xx}$/$\rho_{0}$) at different temperatures with field applied along the $c$-axis. (b) Field dependence of the magnetoresistance (MR) at different temperatures. (c) Phase diagram constructed from the first order derivative of $\rho_{xx}$/$\rho_{0}$ with respect to the magnetic field. The black dashed lines highlight the evolution of the transitions in the $\rho_{xx}$/$\rho_{0}$($H$) curves with temperature.}
\end{figure*}

To understand the overall nature of the $\rho_{xx}$($T$) curve, let us discuss the usual possibilities. The presence of the Kondo effect from the $f$-electrons of cerium could lead to a semiconducting behavior. However, in Kondo materials, $\rho_{xx}$ decreases with decreasing temperature, followed by a region of saturation and then an enhancement at low temperature \cite{Buschow}. In CeSb$_{0.11}$Te$_{1.90}$, in contrast, above $T_{N}$, $\rho_{xx}$ never decreases with decreasing temperature. Moreover, in Ce-based heavy fermionic systems, ARPES measurements with a photon energy above $\sim$100 eV revealed prominent coherent Kondo flat bands at the Fermi level \cite{Fujimori}. From our ARPES results at 110 eV [Fig. 1(b)], we can confirm that such bands are not present in  CeSb$_{0.11}$Te$_{1.90}$ near $E_{F}$. Following the activated transport model of an intrinsic semiconductor, we have plotted $\ln\rho_{xx}$ as a function of 1/$T$ in Fig. S5(a). It is evident that the slope of this curve changes continuously with temperature. Hence, a unique value of any possible CDW-induced energy gap can not be obtained. We note that the $\ln\rho_{xx}$(1/$T$) curves for all field strengths collapse on the zero-field data above $\sim$100 K [Fig. S5(a)]. From the slope of two approximately linear regions, we estimate the energy gap to be $\sim$2.4 and $\sim$0.85 meV with a smooth cross-over between them. Another possible origin of this semiconductor-like behavior can be localization effects due to the intrinsic lattice vacancies ($\delta$) in the single crystals \cite{Lee}. A weak localization effect emerges from quantum interference between scattering paths of carriers at very low-temperature. However, in the case of CeSb$_{0.11}$Te$_{1.90}$, the semiconductor-like transport is very robust up to room temperature. On the other hand, Anderson localization occurs in strongly disordered electron systems at low temperature and gets suppressed under a weak magnetic field. While the resistivity in CeSb$_{0.11}$Te$_{1.90}$ indeed decreases significantly with a small applied field ($<$1 T) around and below $T_{N}$ (probably because of the spin-scattering), the change becomes gradual above $T_{N}$ and $\rho_{xx}$ is found to be insensitive to the magnetic field above $\sim$100 K. Instead, the temperature dependence of the zero-field resistivity is better described by a $\rho_{xx}$=$AT^{-n}$-type power-law, where $A$ and $n$ are arbitrary parameters. Two distinct regions have been observed with the exponent $n\sim$6 for 240$\leq$$T$$\leq$300 K and $n\sim$2 for 60$\leq$$T$$\leq$150 K [Fig. S5(b)]. Such unusual power-law behavior cannot be explained by the conventional scattering models and warrant further investigations. We note that an unconventional power-law dependent resistivity has also been reported for the non-symmorphic Dirac semimetal GdSb$_{0.46}$Te$_{1.48}$, possibly as a result of non-symmorphic Dirac fermions in presence of CDW together with the lattice disorder \cite{Lei2}. However, in the case of GdSb$_{0.46}$Te$_{1.48}$, the obtained exponent $n\sim$0.3.

It is clear that the magnetic field drastically modulates the transport properties of CeSb$_{0.11}$Te$_{1.90}$. In Fig. 3(a), we show the field dependence of the normalized resistivity ($\rho_{xx}$/$\rho_{0}$) at different temperatures, where $\rho_{0}$ is the zero-field resistivity. For both $\rho_{xx}$ and the Hall resistivity ($\rho_{yx}$), a prominent hysteresis is observed between field sweep-up and sweep-down curves. This is a known problem with magnetic materials and requires special attention to avoid any spurious signal during symmetrization or anti-symmetrization of the MR and Hall data, respectively. Therefore, we have adopted a modified symmetrization/anti-symmetrization technique \cite{Czajka2021} to circumvent this issue as described in the supporting information. For $T$$\leq$$T_{N}$, $\rho_{xx}$/$\rho_{0}$ drops sharply in the low field region 0$<$$\mu_{0}H$$\leq$1 T, followed by almost saturation like behavior at higher field strengths. This has some similarities with GMR materials, which are excellent candidates for spin-valve applications \cite{Dieny,Parkin1}. We note that the decrease in resistivity is particularly steep near $T_{N}$, whereas it fades off as we move up/down in temperature. This is consistent with the suppression of the spin-scattering by an external field in magnetic materials. For $T$$>$$T_{N}$, the change in resistivity becomes much more gradual. It is worth mentioning that at low temperatures, $\rho_{xx}$ shows a slight increment in the very narrow field range 0$<$$\mu_{0}H$$\leq$0.03 T [Fig. S6(a)], which could be due to a weak anti-localization effect.

As evident from Fig. 3(a), similar to manganites \cite{Ramirez1,Tokura,Jin}, the magnetic field induced semiconductor-metal-like transition produces a very large negative MR in CeSb$_{0.11}$Te$_{1.90}$. The maximum value of the negative MR, calculated by usual convention, is always restricted to $-$100 \%. Hence, here we use the definition MR=$\frac{\rho_{xx}(H)-\rho_{0}}{\rho_{xx}(H)}\times$100 \%, as commonly used for the manganites \cite{Jin}. At 2 K with a field of 9 T, the MR is found to be $-$1300 \%, which increases to $-$14000 \% at $T_{N}$ [Fig. 3(b)]. To the best of our knowledge, such a large negative CMR has not been observed in TSMs so far and is only one order of magnitude smaller than the ``thousandfold change in resistivity'' in the heat-treated La$_{0.67}$Ca$_{0.33}$MnO$_{x}$ epitaxial film \cite{Jin}. Sample 2, which is a crystal from the same batch and may have an arbitrarily different lattice vacancy, also shows similar behavior with maximum MR=$-$530 \% [Fig. S7(a)]. Furthermore, Murakawa \textit{et al}. recently reported a field-induced sharp decrease in resistivity at the doping range $x$=0.17 for CeSb$_{x}$Te$_{2-x-\delta}$ \cite{Murakawa}. Therefore, the observed CMR is an intrinsic property of CeSb$_{x}$Te$_{2-x-\delta}$ in the heavily electron-doped, multiple \textit{\textbf{q}}-vector regime. Though the MR value diminishes with increasing temperature, it remains noticeable ($-$14 \%) even at 50 K, which is way above $T_{N}$. The MR also possesses a significant anisotropy with respect to the direction of the applied magnetic field. As illustrated in Fig. S8, keeping the current direction unaltered, rotation of the magnetic field in the $ac$-plane generates a two-fold symmetric pattern of the resistivity with $\rho_{xx}$ decreasing much more slowly for fields along the $a$-axis. From the obtained results, we estimate an anisotropy ratio ($\rho_{xx}^{H||a}$/$\rho_{xx}^{H||c}$) $\sim$2.6 at 5.4 K and a magnetic field of 9 T. The large anisotropy ratio is also a confirmation of the layered crystal structure and corresponding quasi two-dimensional nature of the Fermi surface in CeSb$_{0.11}$Te$_{1.90}$.

For $H||c$-axis, below $T_{N}$, the $\rho_{xx}$/$\rho_{0}$ curves show a series of weak transitions within the low-field range 0$<$$\mu_{0}H$$\leq$0.2 T [Fig. S6(a)]. These transitions become readily distinguishable as we take the first order derivative of $\rho_{xx}$/$\rho_{0}$ with respect to the magnetic field [Fig. S6(b)]. In Fig. 3(c), we have constructed a phase diagram by plotting $\partial$($\rho_{xx}$/$\rho_{0}$)/$\partial$($\mu_{0}H$) as a function of temperature and field. As highlighted by the black dashed lines, the temperature evolution of these transitions can be clearly identified. Similar weak transitions are also observed for sample 2 [Fig. S7(b)]. For CeSb$_{x}$Te$_{2-x-\delta}$, only within 0.10$\leq$$x$$<$0.34, CDW modulation wave-vectors extend along the crystallographic $c$-axis \cite{Singha3}. Through RKKY interaction, this CDW gets coupled with the spin modulation, which is created by the alternating spin up/down Ce$^{3+}$ layers along the $c$-axis in the AFM state. With an external magnetic field, the wavelength of the spin modulation can be tuned and as a result these coupled excitations can be driven through successive phase-locked and non-phase-locked states \cite{Singha3}. Therefore, physical properties involving electronic charge density or spins, should both show signatures of these coupled states. In CeSb$_{0.11}$Te$_{1.90}$, similar to the fractionally quantized plateaus in magnetization \cite{Singha3}, the resistivity also shows a string of these transitions. We note that as in magnetization, these transitions disappear in the fully spin-polarized state at $\sim$0.2 T or in the paramagnetic region above $T_{N}$ due to the absence of any spin modulation.

\begin{figure}
\includegraphics[width=0.5\textwidth]{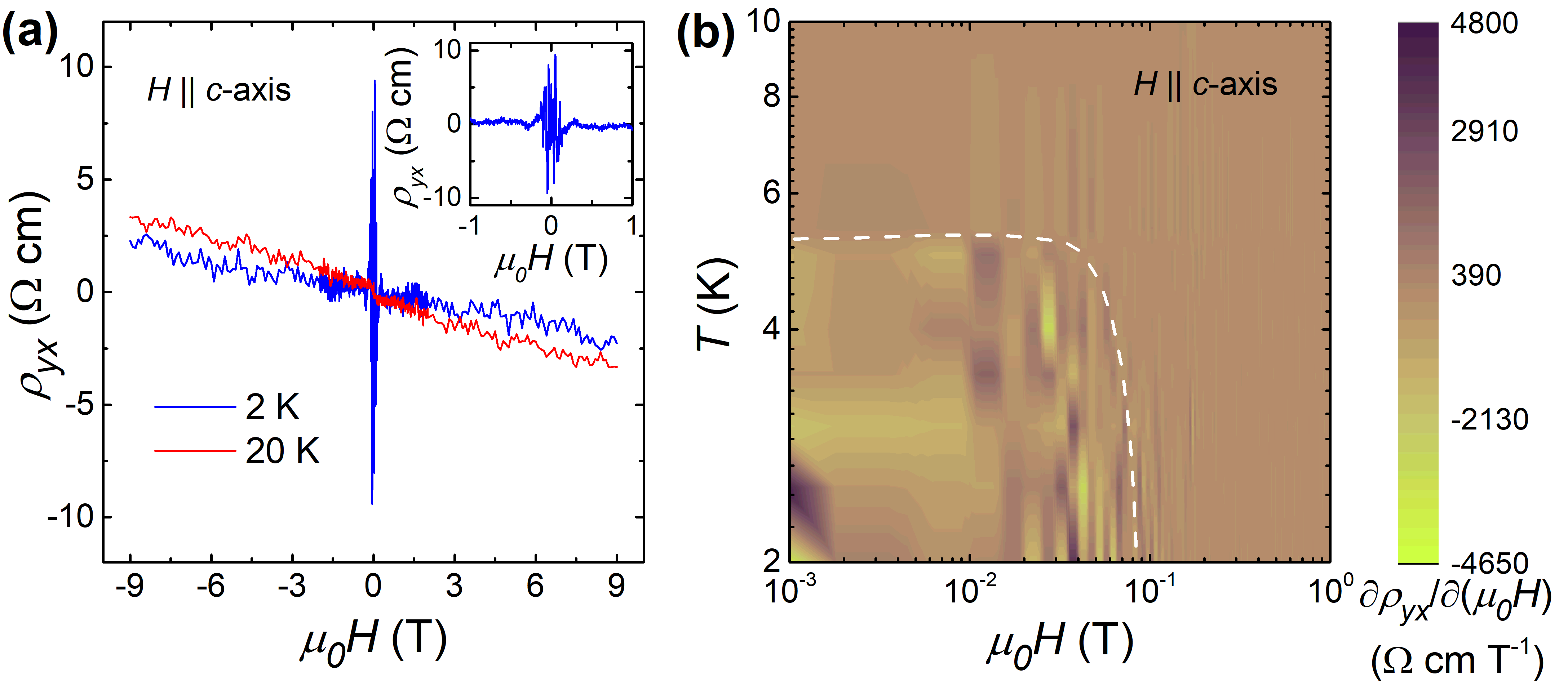}
\caption{Results of the Hall measurements for CeSb$_{0.11}$Te$_{1.90}$. (a) Magnetic field dependence of the Hall resistivity ($\rho_{yx}$) below and above the N{\'e}el temperature. The inset shows the low-field region of $\rho_{yx}$ at 2 K. (b) Phase diagram obtained from the first order derivative of $\rho_{yx}$ with respect to the magnetic field. The white dashed line represents the phase boundary for the anomalous Hall effect.}
\end{figure}

In Fig. 4(a), we have plotted the Hall resistivity as a function of magnetic field at two representative temperatures, above and below $T_{N}$. $\rho_{yx}$($H$) is approximately linear with a negative slope, indicating electron type majority carriers. From the slope in the high-field region at 2 K, we estimate a carrier density $\sim$1.5$\times$10$^{15}$ cm$^{-3}$, which is comparable to intrinsic semiconductors and consistent with the electronic transport properties observed for CeSb$_{0.11}$Te$_{1.90}$. At 2 K, the $\rho_{yx}$($H$) curve shows a prominent discontinuity near the zero-field limit as evident from the inset of Fig. 4(a). We confirm that this giant anomalous Hall effect (AHE) is not a spurious signal coming from the hysteresis during the anti-symmetrization of the experimental data. In Figs. S9(b) and S9(c), we have plotted $\rho_{yx}$($H$), calculated from the conventional method when the hysteresis is neglected. Although it also shows an anomalous behavior, the nature and amplitude is significantly different. In Fig. 4(b), a phase diagram is constructed by plotting $\partial\rho_{yx}$/$\partial$($\mu_{0}H$) as a function of temperature and magnetic field. A phase boundary (indicated by the white line) is obtained which separates the AHE region. We note that this phase boundary almost coincides with the AFM state in magnetic phase diagram. As such an AHE is not expected for a purely AFM ground state with vanishingly small magnetic moment, it could be originated from the collective spin-excitation in the low-field limit or the short range FM ordering.

To summarize, we have probed the high electron filling region of the tunable topological Dirac semimetal CeSb$_{x}$Te$_{2-x-\delta}$ using ARPES and electronic transport measurements. In this doping range, this material is reported to host multiple charge density wave modulation-vectors and a complex magnetic phase diagram. We show that at $x$=0.11, the Dirac semimetal state persists in presence of the antiferromagnetic ordering. Although a metallic Fermi surface is observed, surprisingly, the electronic transport properties resemble semiconductor-like behavior. The overall nature of the resistivity does not match the thermally activated transport or the theoretical localization models. Instead an unusual power-law temperature dependence is identified. An external magnetic field induces a semiconductor-metal-like transition in the low-temperature region. As a consequence, we observe a colossal negative magnetoresistance. The resistivity shows a series of field-induced transitions, which are analogous to the devil's staircase magnetization in CeSb$_{0.11}$Te$_{1.90}$ and originate from the coupling between charge density wave and spin modulation along the crystallographic $c$-axis. This collective spin-excitation possibly also produces a giant anomalous Hall effect in CeSb$_{0.11}$Te$_{1.90}$.

Sample synthesis was supported by an NSF CAREER grant (DMR-2144295) to LMS. ARPES measurements were supported by Air Force office of Scientific Research under award number FA9550-20-1-0246. This work was further supported by the Gordon and Betty Moore Foundation (EPiQS Synthesis Award) through grant GBMF9064 and the David and Lucile Packard Foundation. LMS and NPO are both supported by NSF through the Princeton Center for Complex Materials, a Materials Research Science and Engineering Center DMR-2011750. We thank Klara Volckaert, Davide Curcio, Paulina Majchrzak, and Andreas Topp for their advice on processing of the ARPES data.

\setcounter{figure}{0}

\begin{center}
\textbf{Modified symmetrization and anti-symmetrization of the transport data:}
\end{center}

The experimentally measured magnetoresistance (MR) and Hall data usually contain components from each other due to unavoidable offset in the placement of the electrodes on the sample. This additional component can be excluded by symmetrization (anti-symmetrization) of the MR (Hall) data over positive (pointing up) and negative (pointing down) magnetic field values. For magnetic sample, however, this method can generate spurious signals as there can be hysteresis between field sweep-up (from negative to positive) and sweep-down (from positive to negative) curves. For example, in CeSb$_{0.11}$Te$_{1.90}$, the measured Hall resistivity ($\rho_{yx}^{raw}$) at 2 K in Fig. S9(a), demonstrate clear hysteresis between these two curves. To tackle this problem, we measured the resistivity ($\rho_{xx}^{raw}$) and Hall resistivity for both field sweep up ($H_{\uparrow}$) and sweep down ($H_{\downarrow}$) at each temperature. The actual $\rho_{xx}$ and $\rho_{yx}$ are then calculated using the following equations \cite{Czajka2021},

\begin{equation}
    \rho_{xx}(H)=\frac{1}{2} [\rho_{xx}^{raw}(H_{\uparrow})+\rho_{xx}^{raw}(-H_{\downarrow})]
\end{equation}

\begin{equation}
    \rho_{xx}(-H)=\frac{1}{2} [\rho_{xx}^{raw}(-H_{\downarrow})+\rho_{xx}^{raw}(H_{\uparrow})]
\end{equation}

\begin{equation}
    \rho_{yx}(H)=\frac{1}{2} [\rho_{yx}^{raw}(H_{\uparrow})-\rho_{yx}^{raw}(-H_{\downarrow})]
\end{equation}

\begin{equation}
    \rho_{yx}(-H)=\frac{1}{2} [\rho_{yx}^{raw}(-H_{\downarrow})-\rho_{yx}^{raw}(H_{\uparrow})]
\end{equation}

\begin{figure*}
\includegraphics[width=0.8\textwidth]{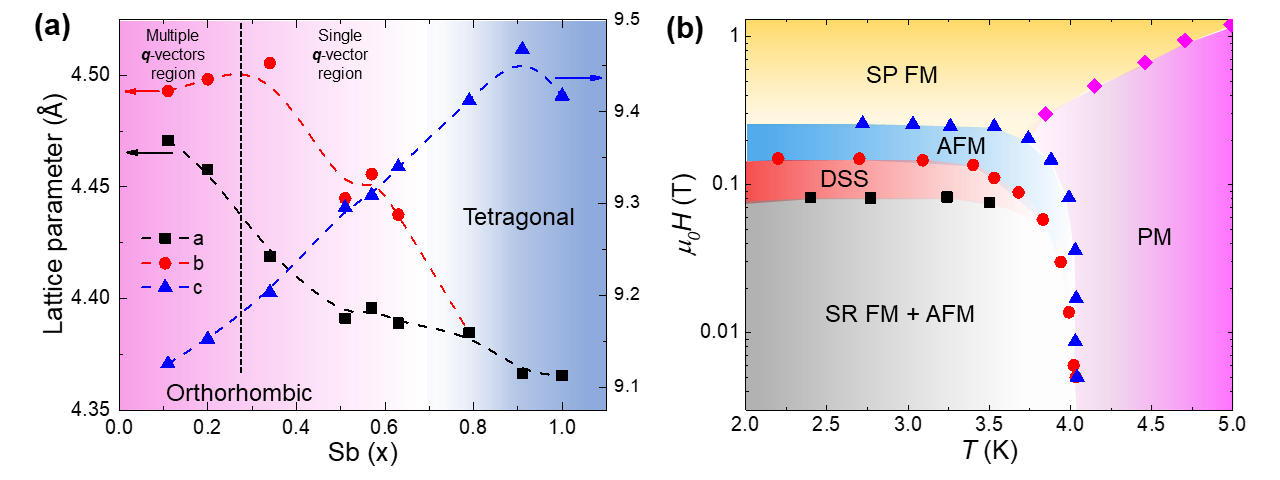}
\renewcommand{\figurename}{Fig. S}
\caption{Phase diagram of CeSb$_{x}$Te$_{2-x-\delta}$ reproduced from the data in Ref. \cite{Singha2021}. (a) Structural phase diagram constructed from lattice parameters with changing electron filling at the Sb square-net. (b) Magnetic phase diagram for CeSb$_{0.11}$Te$_{1.90}$. The phase boundaries separate the antiferromagnetic (AFM), short-range ferromagnetic (SR FM), spin-polarized ferromagnetic (SP FM), paramagnetic (PM), and devil's staircase (DSS) states.}
\end{figure*}

\begin{figure*}
\includegraphics[width=0.8\textwidth]{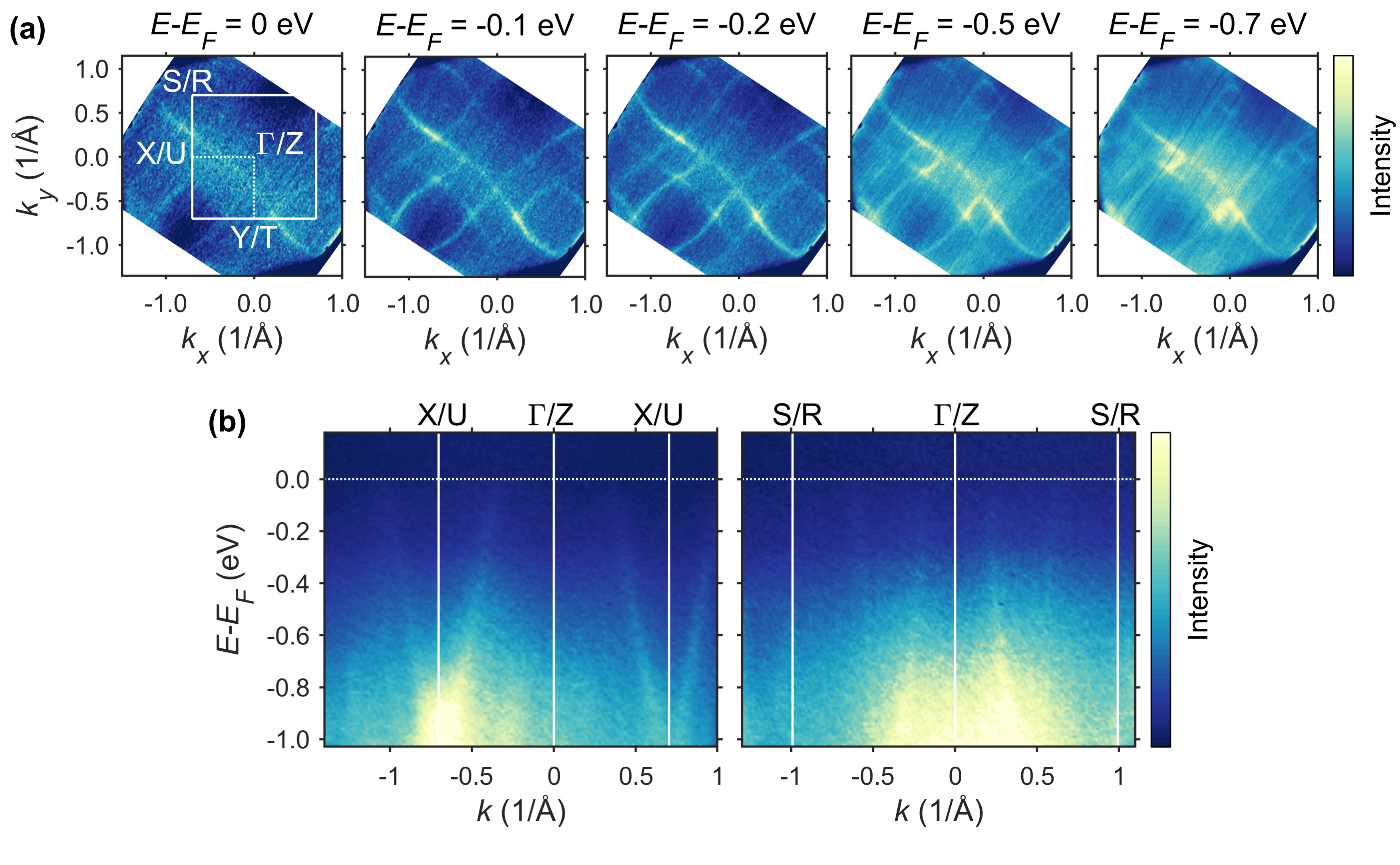}
\renewcommand{\figurename}{Fig. S}
\caption{ARPES spectra for CeSb$_{0.11}$Te$_{1.90}$ at $\sim$30 K measured with a photon energy of 70 eV. (a) Constant energy contours at different energy ($E$) values from Fermi energy ($E_{F}$) to $E_{F}-$0.7 eV. (b) Electronic band dispersion along the high symmetry directions $X$($U$)-$\Gamma$($Z$)-$X$($U$) and $S$($R$)-$\Gamma$($Z$)-$S$($R$).}
\end{figure*}

\begin{figure*}
\includegraphics[width=0.9\textwidth]{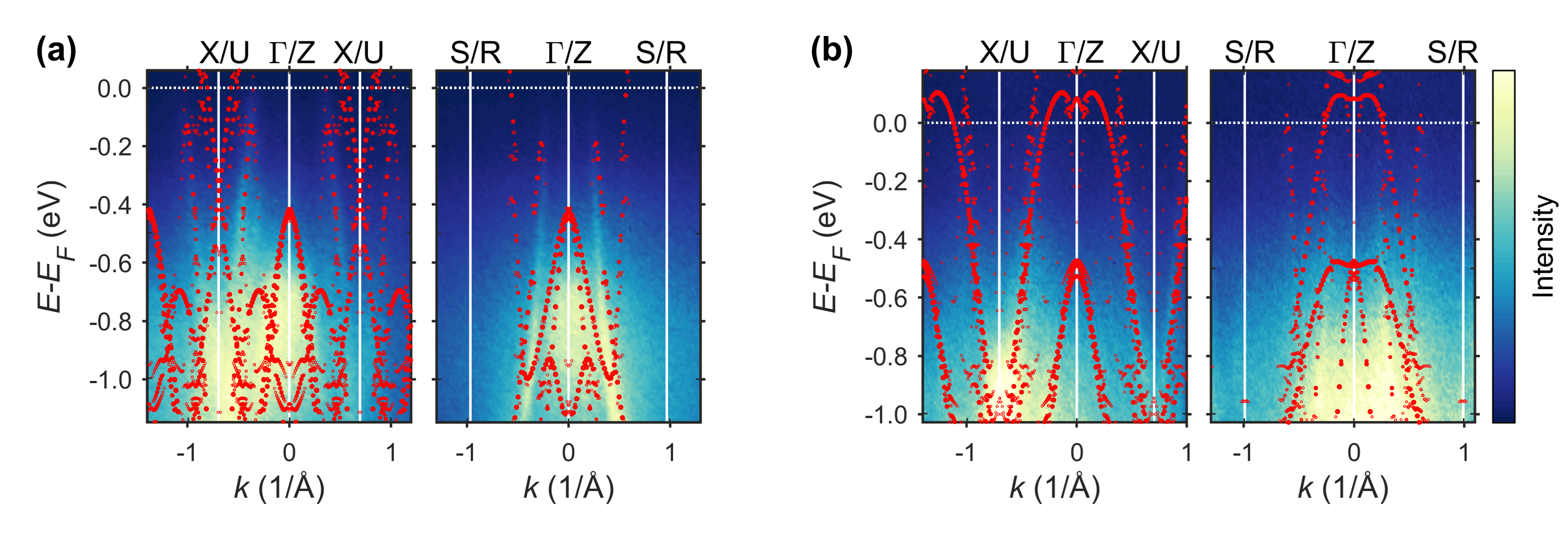}
\renewcommand{\figurename}{Fig. S}
\caption{Overlay of the theoretical band structure with the ARPES data for photon energy (a) 110 eV and (b) 70 eV. The calculated band structure is plotted with red circles along the $k$-path $U$-$Z$-$U$ and $R$-$Z$-$R$ in (a) and $X$-$\Gamma$-$X$ and $S$-$\Gamma$-$S$ in (b). Calculated points with weightage less than 4 \% of the maximum of the path were excluded from the plot.}
\end{figure*}

\begin{figure*}
\includegraphics[width=0.4\textwidth]{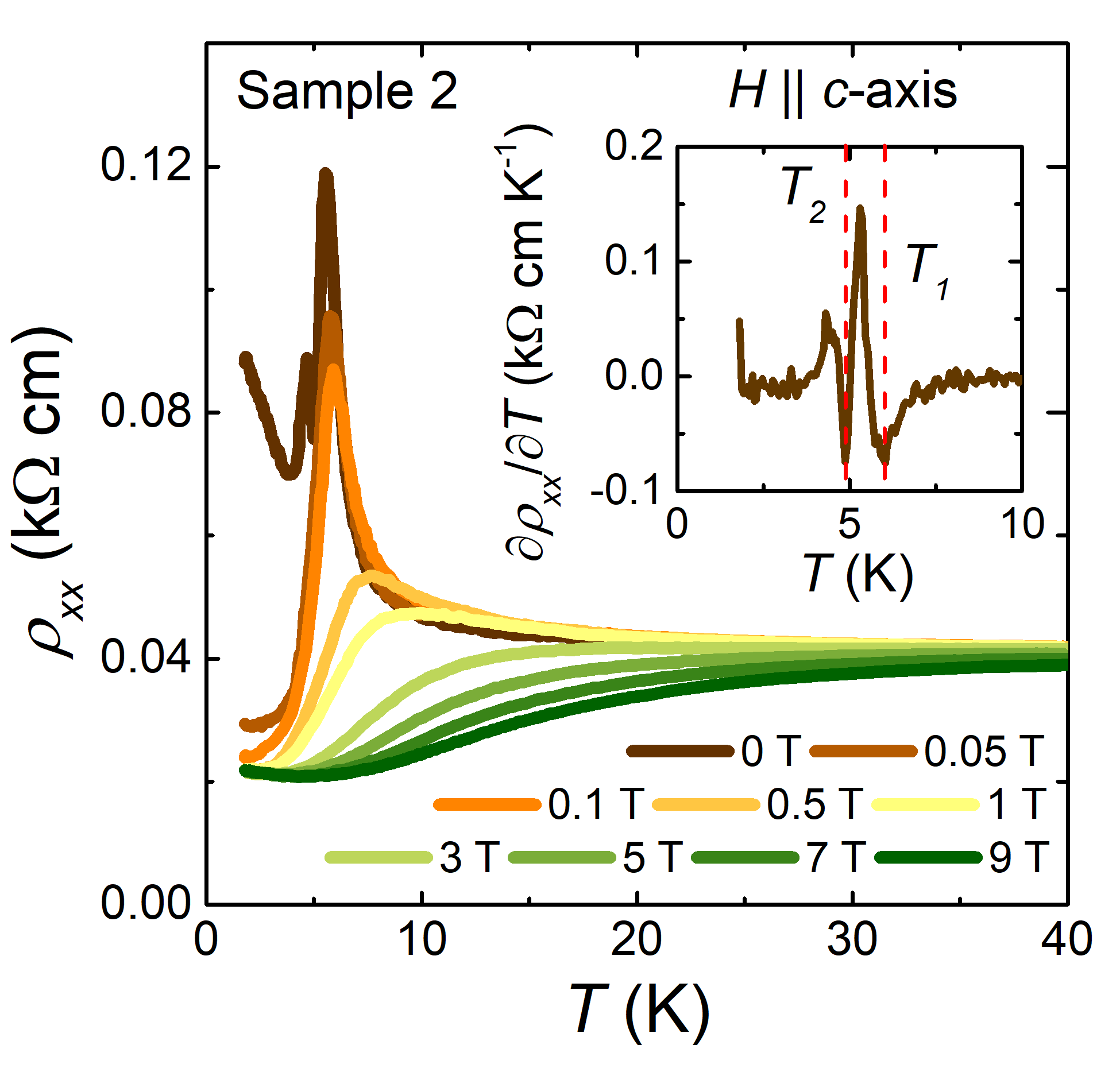}
\renewcommand{\figurename}{Fig. S}
\caption{Temperature dependent electronic transport properties for the second crystal (sample 2) from the same batch. Temperature ($T$) dependence of the resistivity ($\rho_{xx}$) at different external magnetic fields applied along the $c$-axis. The first order derivative of $\rho_{xx}$($T$) is plotted in the inset showing the transition temperatures (vertical dashed lines).}
\end{figure*}

\begin{figure*}
\includegraphics[width=0.7\textwidth]{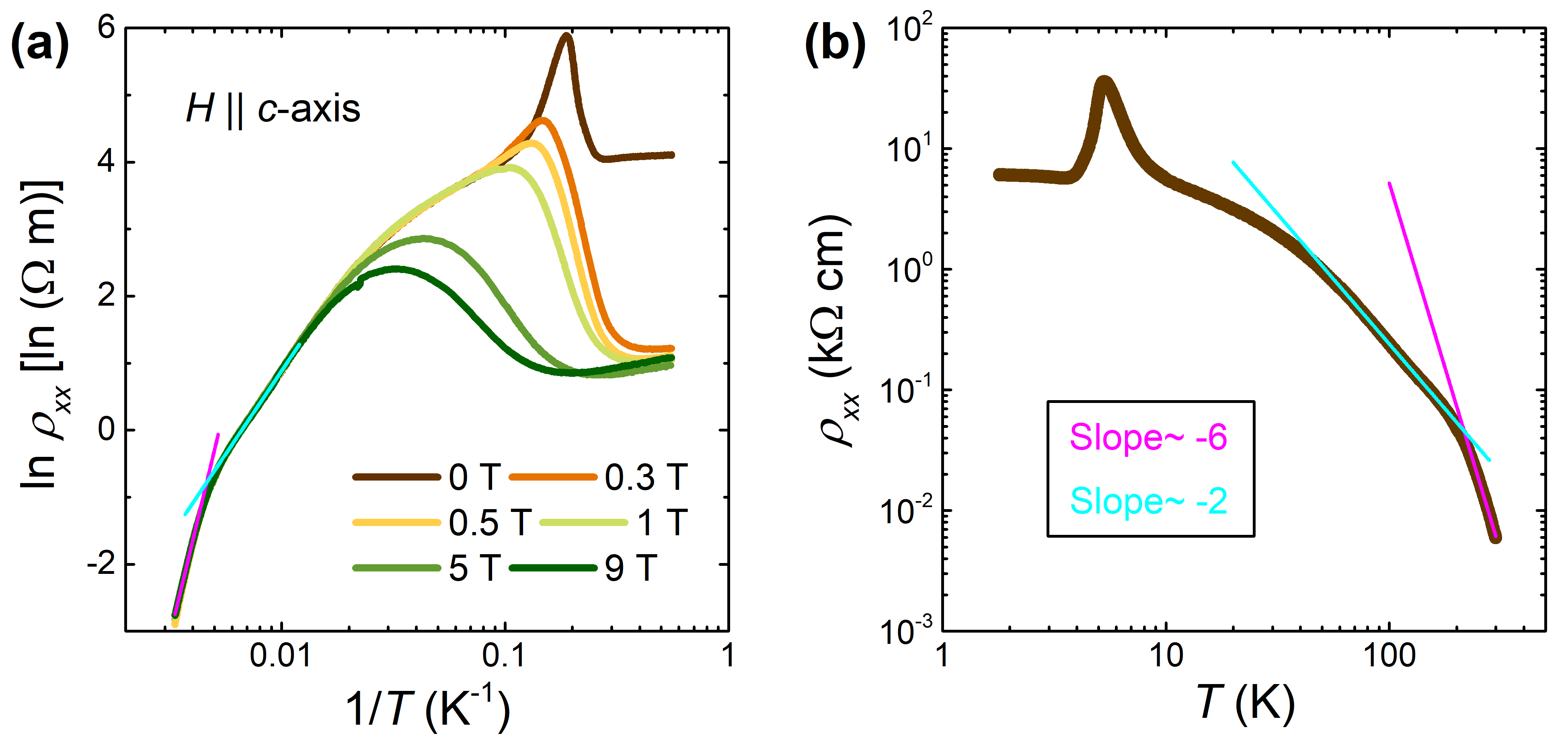}
\renewcommand{\figurename}{Fig. S}
\caption{Fitting of the temperature dependent resistivity for CeSb$_{0.11}$Te$_{1.90}$. (a) Natural logarithmic resistivity as a function of inverse temperature. The magenta and cyan lines represent fitting of two approximately linear regions. (b) Power-law temperature dependence of the resistivity. The linear fittings (magenta and cyan lines) illustrate two different exponents.}
\end{figure*}

\begin{figure*}
\includegraphics[width=0.7\textwidth]{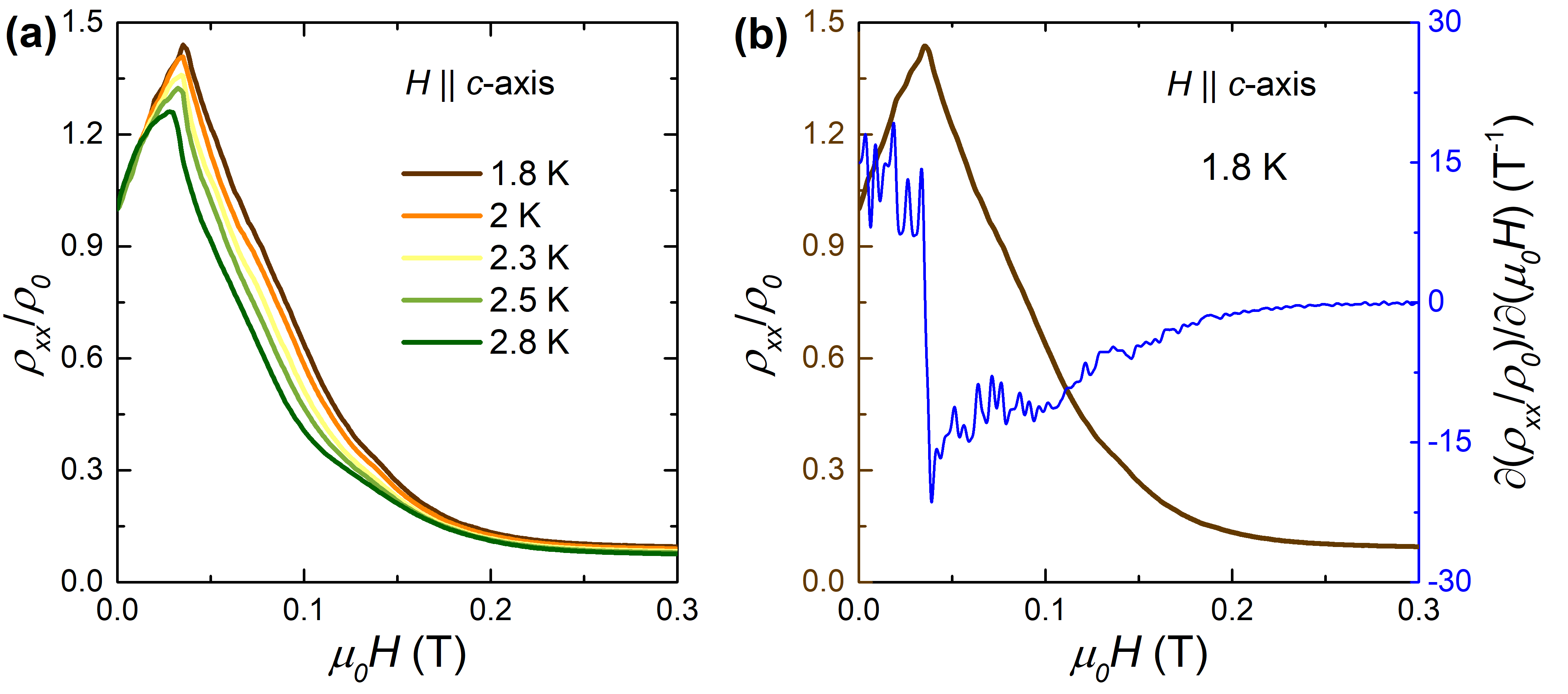}
\renewcommand{\figurename}{Fig. S}
\caption{Low magnetic-field region of the normalized resistivity for CeSb$_{0.11}$Te$_{1.90}$. (a) Low magnetic-field region of the normalized resistivity ($\rho_{xx}$/$\rho_{0}$) at different temperatures with the field applied along the $c$-axis. A series of transitions can be readily identified. (b) The first order derivative of $\rho_{xx}$/$\rho_{0}$ at a representative temperature 1.8 K to clearly resolve the weak transitions.}
\end{figure*}

\begin{figure*}
\includegraphics[width=0.75\textwidth]{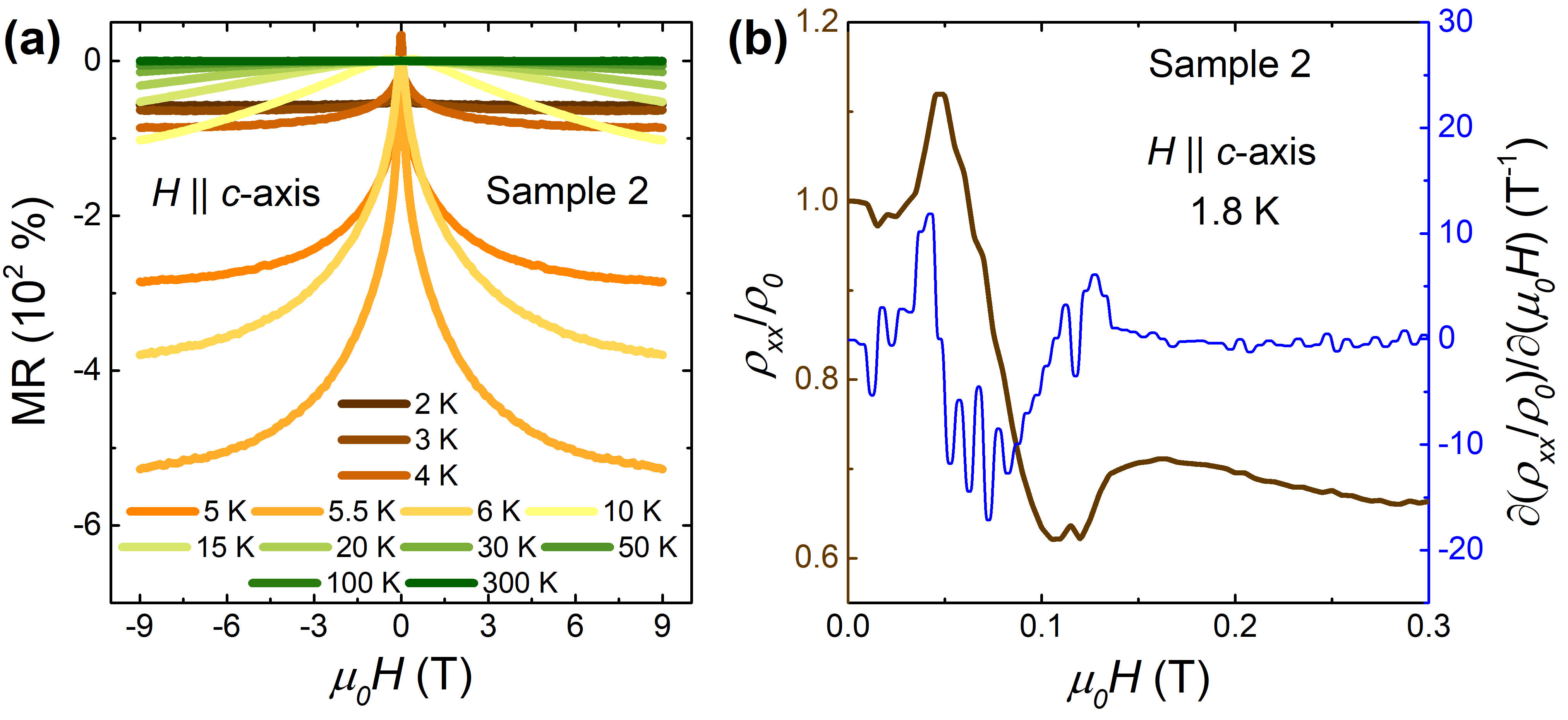}
\renewcommand{\figurename}{Fig. S}
\caption{Magnetic field dependent electronic transport properties of sample 2. (a) Magnetoresistance at different temperatures for magnetic field applied along the $c$-axis. (b) Low-field region of the $\rho_{xx}$/$\rho_{0}$ curve along with the first order derivative with respect to the magnetic field at 1.8 K.}
\end{figure*}

\begin{figure*}
\includegraphics[width=0.7\textwidth]{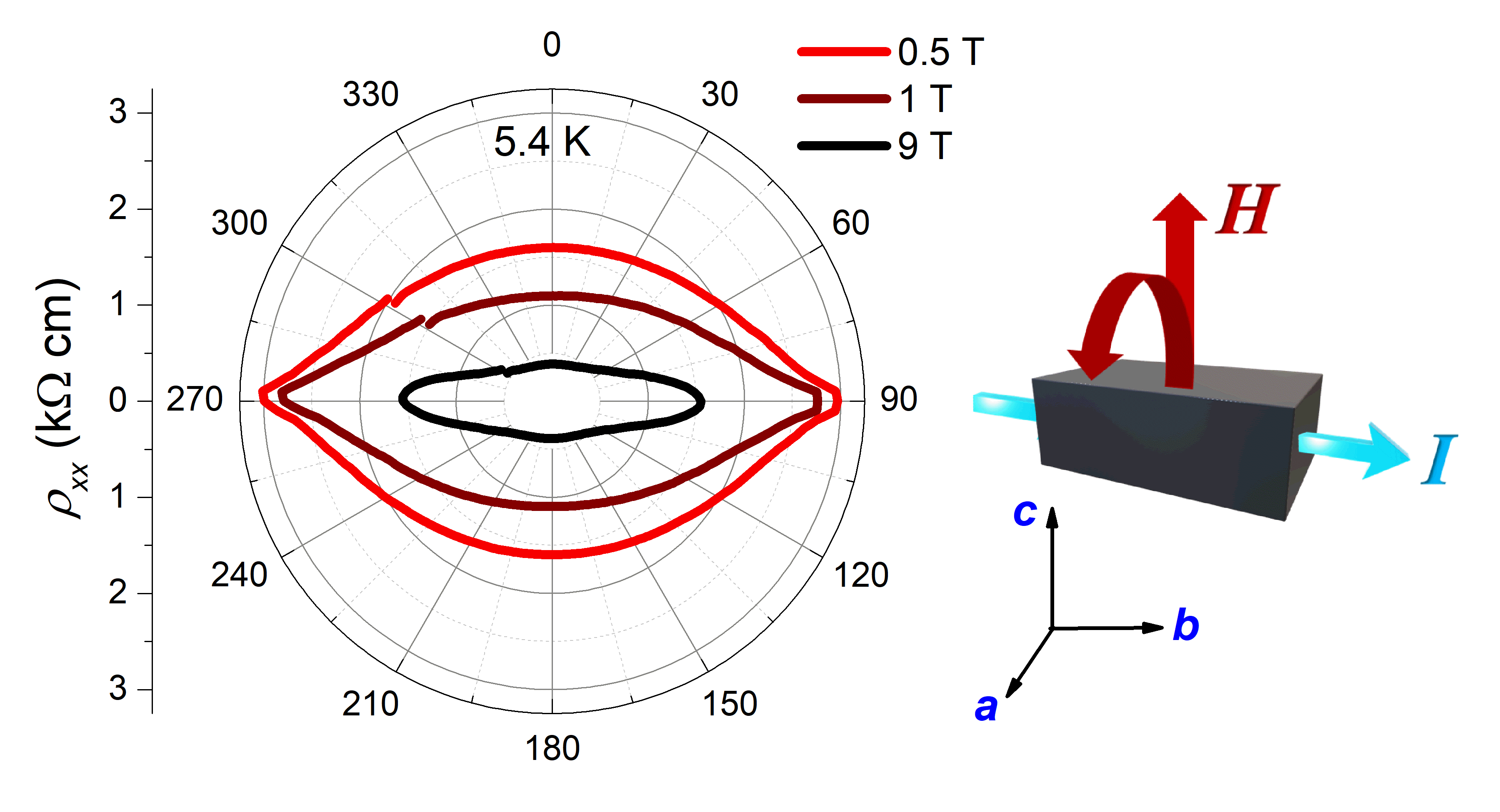}
\renewcommand{\figurename}{Fig. S}
\caption{Anisotropy of the magnetic field-dependent resistivity for CeSb$_{0.11}$Te$_{1.90}$. Crystallographic direction dependence of the resistivity at 5.4 K with different magnetic field strengths, when the current is along the $b$-axis and field is rotated in the $ac$-plane. The schematic illustrates the measurement configuration.}
\end{figure*}

\begin{figure*}
\includegraphics[width=1.0\textwidth]{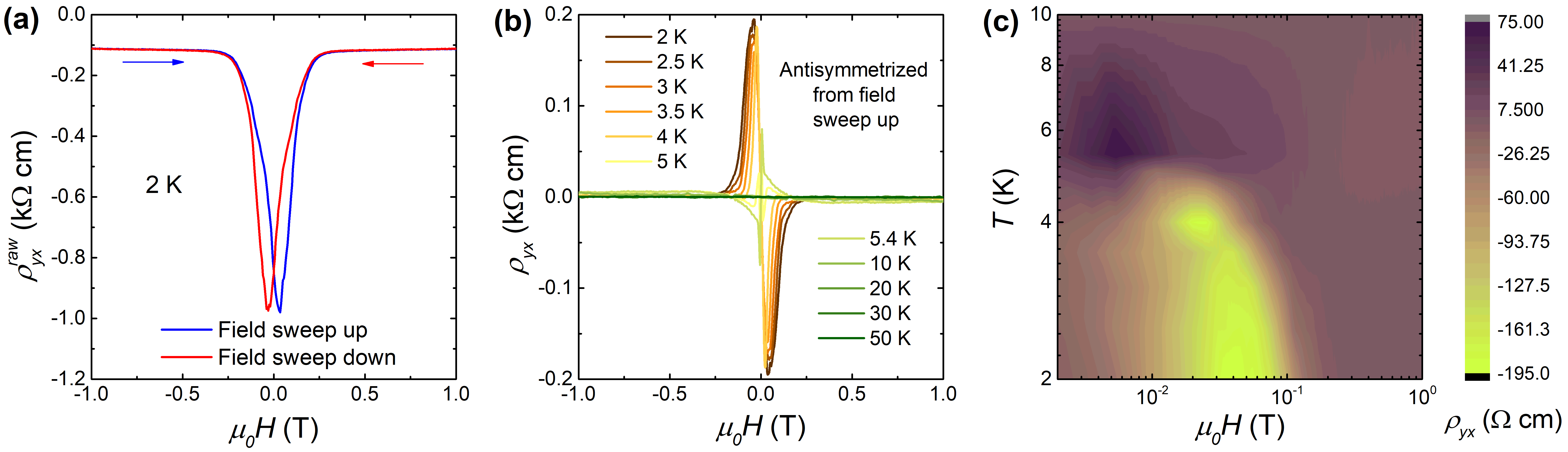}
\renewcommand{\figurename}{Fig. S}
\caption{Modified anti-symmetrization of the Hall data for antiferromagnetic CeSb$_{0.11}$Te$_{1.90}$. (a) Hysteresis between the magnetic field sweep-up (-9 to 9 T) and sweep-down (9 to -9 T) curves of the raw Hall resistivity ($\rho_{yx}^{raw}$) data. (b) Extracted Hall resistivity ($\rho_{yx}$) at different temperatures, obtained by anti-symmetrizing the raw data using a single field sweep. (c) Phase diagram of $\rho_{yx}$, calculated from a single field sweep.}
\end{figure*}

\end{document}